\documentclass[longauth,letter]{aa}
\usepackage{graphicx}
\usepackage{lscape}
\usepackage{txfonts}

\def\msol{\hbox{\kern 0.20em $M_\odot$}}

\newcommand{\lsol}{\hbox{\kern 0.20em $L_\odot$}}
\newcommand{\g}{\hbox{\kern 0.20em g}}
\newcommand{\gmu}{\hbox{\kern 0.20em g$^{-1}$}}
\newcommand{\kg}{\hbox{\kern 0.20em kg}}
\newcommand{\pc}{\hbox{\kern 0.20em pc}}
\newcommand{\mum}{\hbox{\kern 0.20em $\mu$m}}
\newcommand{\mumd}{\hbox{\kern 0.20em $\mu$m$^{-2}$}}
\newcommand{\cm}{\hbox{\kern 0.20em cm}}
\newcommand{\m}{\hbox{\kern 0.20em m}}
\newcommand{\km}{\hbox{\kern 0.20em km}}
\newcommand{\nm}{\hbox{\kern 0.20em nm}}
\newcommand{\s}{\hbox{\kern 0.20em s}}
\newcommand{\h}{\hbox{\kern 0.20em h}}
\newcommand{\smu}{\hbox{\kern 0.20em s$^{-1}$}}
\newcommand{\srmu}{\hbox{\kern 0.20em sr$^{-1}$}}
\newcommand{\smd}{\hbox{\kern 0.20em s$^{-2}$}}
\newcommand{\an}{\hbox{\kern 0.20em an}}
\newcommand{\anmu}{\hbox{\kern 0.20em an$^{-1}$}}
\newcommand{\yr}{\hbox{\kern 0.20em yr}}
\newcommand{\yrmu}{\hbox{\kern 0.20em yr$^{-1}$}}
\newcommand{\Myr}{\hbox{\kern 0.20em Myr}}
\newcommand{\Mymu}{\hbox{\kern 0.20em Myr$^{-1}$}}
\newcommand{\K}{\hbox{\kern 0.20em K}}
\newcommand{\pcmu}{\hbox{\kern 0.20em pc$^{-1}$}}
\newcommand{\pcmd}{\hbox{\kern 0.20em pc$^{-2}$}}
\newcommand{\pcmt}{\hbox{\kern 0.20em pc$^{-3}$}}
\newcommand{\kms}{\hbox{\kern 0.20em km\kern 0.20em s$^{-1}$}}
\newcommand{\kmpd}{\hbox{\kern 0.20em km$^{2}$}}
\newcommand{\kpc}{\hbox{\kern 0.20em kpc}}
\newcommand{\cms}{\hbox{\kern 0.20em cm\kern 0.20em s$^{-1}$}}
\newcommand{\erg}{\hbox{\kern 0.20em erg}}
\newcommand{\ergs}{\hbox{\kern 0.20em erg}}
\newcommand{\cmpd}{\hbox{\kern 0.20em cm$^2$}}
\newcommand{\cmmd}{\hbox{\kern 0.20em cm$^{-2}$}}
\newcommand{\cmms}{\hbox{\kern 0.20em cm$^{-6}$}}
\newcommand{\cmpt}{\hbox{\kern 0.20em cm$^3$}}
\newcommand{\cmmt}{\hbox{\kern 0.20em cm$^{-3}$}}
\newcommand{\mpd}{\hbox{\kern 0.20em m$^2$}}
\newcommand{\mmd}{\hbox{\kern 0.20em m$^{-2}$}}
\newcommand{\mpt}{\hbox{\kern 0.20em m$^3$}}
\newcommand{\mmt}{\hbox{\kern 0.20em m$^{-3}$}}
\newcommand{\mujy}{\hbox{\kern 0.20em $\mu$Jy}}
\newcommand{\mjy}{\hbox{\kern 0.20em mJy}}
\newcommand{\Mj}{\hbox{\kern 0.20em MJy}}
\newcommand{\jy}{\hbox{\kern 0.20em Jy}}
\newcommand{\ghz}{\hbox{\kern 0.20em GHz}}
\newcommand{\G}{\hbox{\kern 0.20em G}}
\newcommand{\muG}{\hbox{\kern 0.20em $\mu$G}}

\newcommand{\cs}{\hbox{CS}}

\newcommand{\htwo}{\hbox{H${}_2$}}

\newcommand{\water}{\hbox{H$_{2}$O}}
\newcommand{\gwater}{\hbox{o-H$_{2}$O {}$1_{10}-1_{01}$}}

\begin{document}
\title{CHESS: Chemical Herschel surveys of star forming regions~:
Peering into the protostellar shock L1157-B1.
\thanks{{\em Herschel} is an ESA space observatory with science instruments provided
by European-led Principal Investigator consortia and with important
participation from NASA.}} \subtitle {II.~Shock dynamics}

\author{
%
%
Lefloch B. \inst{1}  \and Cabrit S. \inst{2}  \and Codella C.\inst{3} \and
Melnick G. \inst{4} \and Cernicharo J.\inst{5}  \and Caux E. \inst{6}   \and
Benedettini M. \inst{7} \and Boogert A. \inst{8} \and Caselli P. \inst{9} \and
Ceccarelli C. \inst{1} \and Gueth F. \inst{10} \and Hily-Blant P. \inst{1} \and
Lorenzani A. \inst{3} \and Neufeld D. \inst{11} \and Nisini B. \inst{12} \and
Pacheco S. \inst{1} \and Pagani L. \inst{2} \and Pardo J.R. \inst{5} \and
Parise B. \inst{13} \and Salez M. \inst{2} \and Schuster K. \inst{10} \and Viti
S. \inst{12,14} \and
%
%
Bacmann A.\inst{1,15} \and Baudry A. \inst{15} \and Bell T. \inst{16} \and
Bergin E.A. \inst{17} Blake G. \inst{16} \and Bottinelli S. \inst{6} \and
Castets A. \inst{1} \and Comito C. \inst{13} \and Coutens A. \inst{6} \and
Crimier N. \inst{1,5} \and Dominik C. \inst{18,29} \and Demyk K. \inst{6} \and
Encrenaz P. \inst{2} \and Falgarone E. \inst{2} \and Fuente A. \inst{20} \and
Gerin M. \inst{2} \and Goldsmith P. \inst{21} \and Helmich F. \inst{22} \and
Hennebelle P.\inst{2} \and Henning T. \inst{23} \and Herbst E. \inst{24} \and
Jacq T. \inst{15} \and Kahane C. \inst{1} \and Kama M. \inst{18} \and Klotz A.
\inst{6} \and Langer W. \inst{21} \and Lis D. \inst{16} \and Lord S. \inst{16}
\and Maret S. \inst{1} \and Pearson J. \inst{21} \and Phillips T. \inst{16}
\and Saraceno P. \inst{7} \and Schilke P. \inst{13,25} \and Tielens A.G.G.M.
\inst{26} \and van der Tak F. \inst{22,19} \and van der Wiel M. \inst{19,22}
\and Vastel C. \inst{6} \and Wakelam V. \inst{15} \and Walters A. \inst{6} \and
Wyrowski F. \inst{13} \and Yorke H. \inst{21}
%
%
\and Bachiller R. \inst{20} \and Borys C. \inst{16} \and De Lange G. \inst{22}
\and Delorme Y. \inst{5} \and Kramer C. \inst{25,27} \and Larsson B. \inst{28}
\and Lai R. \inst{30} \and Maiwald F.W. \inst{21} \and Martin-Pintado J.
\inst{5} \and Mehdi I. \inst{21} \and Ossenkopf V. \inst{25} \and Siegel P.
\inst{21} \and Stutzki J. \inst{25} \and Wunsch J.H.\inst{13} }

\institute{
Laboratoire d'Astrophysique de Grenoble, UMR 5571-CNRS, Universit\'e Joseph
Fourier, Grenoble: \email{lefloch@obs.ujf-grenoble.fr} \and
Observatoire de Paris-Meudon, LERMA UMR CNRS 8112. Meudon, France \and
INAF, Osservatorio Astrofisico di Arcetri, Firenze, Italy \and
Center for Astrophysics, Cambridge MA, USA \and
Centro de Astrobiolog\'{\i}a, CSIC-INTA, Madrid, Spain \and
CESR, Universit\'e Toulouse 3 and CNRS, Toulouse, France \and
INAF, Istituto di Fisica dello Spazio Interplanetario, Roma, Italy \and
Infared Processing and Analysis Center, Caltech, Pasadena, USA \and
School of Physics and Astronomy, University of Leeds, Leeds, UK \and
Institut de Radio Astronomie Millim\'etrique, Grenoble, France \and
Johns Hopkins University, Baltimore MD, USA \and
INAF, Osservatorio Astronomico di Roma, Monte Porzio Catone, Italy \and
Max-Planck-Institut f\"{u}r Radioastronomie, Bonn, Germany \and
Department of Physics and Astronomy, University College London, London, UK \and
Universit\'{e} de Bordeaux, Laboratoire d'Astrophysique de Bordeaux, France;
CNRS/INSU, Floirac, France \and
California Institute of Technology, Pasadena, USA \and
University of Michigan, Ann Arbor, USA \and
Astronomical Institute 'Anton Pannekoek', University of Amsterdam, Amsterdam,
The Netherlands \and
Kapteyn Astronomical Institute, University of Groningen, Groningen, The
Netherlands \and
IGN Observatorio Astron\'{o}mico Nacional, Spain \and
Jet Propulsion Laboratory, Caltech, Pasadena, CA 91109, USA \and
SRON, Groningen, The Netherlands \and
Max Planck Institut f\"ur Astronomie , Heidelberg - Germany \and
Ohio State University, Columbus, OH, USA \and
Physikalisches Institut, Universit\"{a}t zu K\"{o}ln, K\"{o}ln, Germany \and
Leiden Observatory, Leiden University, Leiden, The Netherlands \and
IRAM, Granada, Spain \and
Department of Astronomy, Stockholm University, Stockholm, Sweden \and
Radboud University Nijmegen, The Netherlands \and
Northrop Grumman Aerospace Systems, Redondo Beach, CA 90278 U.S.A }

\date{2010 March 31; 2010 April 30}
\abstract{The outflow driven by the low-mass class 0 protostar L1157 is the
prototype of the so-called chemically active outflows. The bright bowshock B1
in the southern outflow lobe is a privileged testbed of magneto-hydrodynamical
(MHD) shock models, for which dynamical and chemical processes are strongly
interdependent.}{We present the first results of the unbiased spectral survey
of the L1157-B1 bowshock, obtained in the framework of the key program
"Chemical Herschel Surveys of Star Forming Regions" (CHESS). The main aim is to
trace the warm and chemically enriched gas  and to infer the excitation
conditions in the shock region.} {The CO 5-4 and \gwater\ lines have been
detected at high-spectral resolution in the unbiased spectral survey of the
HIFI-Band 1b spectral window (555-636 GHz), presented by Codella et al. in this
volume. Complementary ground-based observations in the submm window help
establish the origin of the emission detected in the main-beam of HIFI, and the
physical conditions in the shock.}{Both lines exhibit broad wings, which extend
to velocities much higher than reported up to now.  We find that the molecular
emission arises from two regions with distinct physical conditions~: an
extended, warm ($100\K$), dense ($3\times 10^5\cmmt$) component at
low-velocity, which dominates the water line flux in Band~1; a secondary
component in a small region of B1 (a few arcsec) associated with high-velocity,
hot ($ > 400\K$) gas of  moderate density ($(1.0-3.0)\times 10^4\cmmt$), which
appears to dominate the flux of the water line at $179\mum$ observed with PACS.
The water abundance is enhanced by two orders of magnitude between the low- and
the high-velocity component, from $8\times 10^{-7}$ up to $8\times 10^{-5}$.
The properties of the high-velocity component agree well with the predictions
of steady-state C-shock models.}{}

\titlerunning{CHESS spectral survey of L1157-B1~: Shock Dynamics}
\authorrunning{Lefloch et al.}
\maketitle

\section{Introduction}

Shocks in protostellar outflows play a crucial role in the  molecular cloud
evolution and star formation by transferring momentum and energy back to the
ambient medium. There is mounting evidence that these shocks often involve a
magnetic precursor where ionic and neutral species are kinematically decoupled.
Magneto-hydrodynamical (MHD) shocks are important not only for the cloud
dynamics, but also for the chemical evolution through temperature and density
changes, which favors the activation of endothermic reactions, ionization, and
dust destruction through sputtering and shattering in the ion neutral drift
zone.  These various processes lead to abundance enhancements up to several
orders of magnitude, as reported for various molecular species in "chemically
active" outflows (Bachiller et al. 2001).
Conversely, the magnetic field and the ionization fraction play an important
role in controlling the size and the temperature of the ion-neutral drift zone.
Because of the interplay between the dynamics and chemistry, the physics of MHD
shocks requires a comprehensive picture of both the gas and dust physical
conditions in the compressed region itself.

Along with $\htwo$, $\water$ and CO are two key-molecules predicted to dominate
the cooling of MHD shocks (Kaufman \& Neufeld 1996). The abundance of \water\
in protostellar regions can be greatly enhanced in shocks, even of moderate
velocity. This occurs both from the sputtering of frozen water from grain
mantles and through high-temperature sensitive reactions in the gas phase
(Elitzur \& de Jong, 1978; Elitzur \& Watson, 1978; Bergin et al. 1998).
Multi-transition observations of  these two molecules therefore serve as good
probes of shock regions with  various excitation conditions, and can be used to
set stringent constraints on MHD shock models (Flower \& Pineau des Forets,
2010).


The heterodyne instrument HIFI onboard {\em Herschel} allows us to study with
unprecedented sensitivity the chemical and dynamical evolution of protostellar
shocks, at spectral and angular resolutions comparable to the best ground-based
single-dish telescopes. This is the main goal of the spectral survey of
L1157-B1, carried out in the guaranteed time key-project CHESS.

The source L1157-mm is a low-mass Class 0 protostar located at a distance
estimated between 250~pc (Looney et al. 1997) and 440~pc (Viotti 1969). It
drives a spectacular bipolar outflow, which has been studied in detail at
millimeter and far-infrared wavelengths.
Mapping of the southern  lobe of L1157  with the Plateau de Bure Interferometer
(PdBI) reveals two limb-brightened cavities (Gueth et al. 1996), each one
terminated by a strong bow shock, dubbed "B1" and "B2" respectively (Fig.~1),
which are likely the result of episodic ejection events in a precessing, highly
collimated jet. The spatial and kinematical structure of B1 has been modelled
in great detail by various authors, making it the archetype of protostellar
bowshocks in low-mass star-forming regions and the testbed of MHD shock models
(Gusdorf et al. 2008a,b).

Here, we report on the emission lines of CO and \water\ detected in the
low-frequency band of HIFI in the course of the CHESS spectral survey. From
comparison with complementary observations, we discuss the origin of the
emission, and based on a simple modelling of the source, we derive the water
abundance in the shock region.

\section{Observations and results}

A full coverage of the band 1b at $20^h39^m10.2^s$ $+68^{\circ}01'10.5\arcsec$
(J2000) in the bowshock B1 was carried out with the HIFI heterodyne instrument
(de Graauw et al. 2010) on board of the {\em Herschel} Space Observatory
(Pilbratt et al. 2010) during the Performance Verification phase on 2009 August
1. The corresponding dataset is OBS$\_$1342181160. The HIFI band 1b (from 555.4
to 636.2 GHz) was covered in double beam switching. Both polarizations (H and
V) were observed simultaneously. The receiver was tuned in double sideband
(DSB) with a total integration time of 140 minutes. In order to obtain the best
possible data reconstruction, the survey was acquired with a degree of
redundancy of 4. The Wide Band Spectrometer (WBS) was used as spectrometer,
providing a frequency resolution of 0.5~MHz.

The data were processed with the ESA-supported package HIPE (Ott et al. 2009).
Fits file from level 2 were then created and transformed into GILDAS
format\footnote{http://www.iram.fr/IRAMFR/GILDAS} for baseline subtraction and
subsequent sideband deconvolution. The spectral resolution was degraded to 1MHz
in the final single sideband (SSB) dataset. The calibration for each receiver
(H and V) is better than 2-3\%. The relative calibration between both receivers
is also rather good, with a difference in intensity of about 4\%. The overall
calibration uncertainty is about $7\%$, except for the strong CO line present
in the band (see below).

Two strong lines dominate over the molecular transitions detected in the
spectral band: the fundamental line of water in its ortho state \gwater\ at
556.936069~GHz and the CO 5-4 line at 576.276905~GHz (Fig.~1). The final rms
noise is 13 mK.
We adopted the theoretical telescope main-beam efficiency $\eta_{mb}= 0.723$,
and a main-beam size of $39\arcsec$ (HPFW) in the whole band. Unless indicated,
intensities are expressed in units of antenna temperature $\rm T_{A}$.

\begin{figure}
\includegraphics[width=\columnwidth]{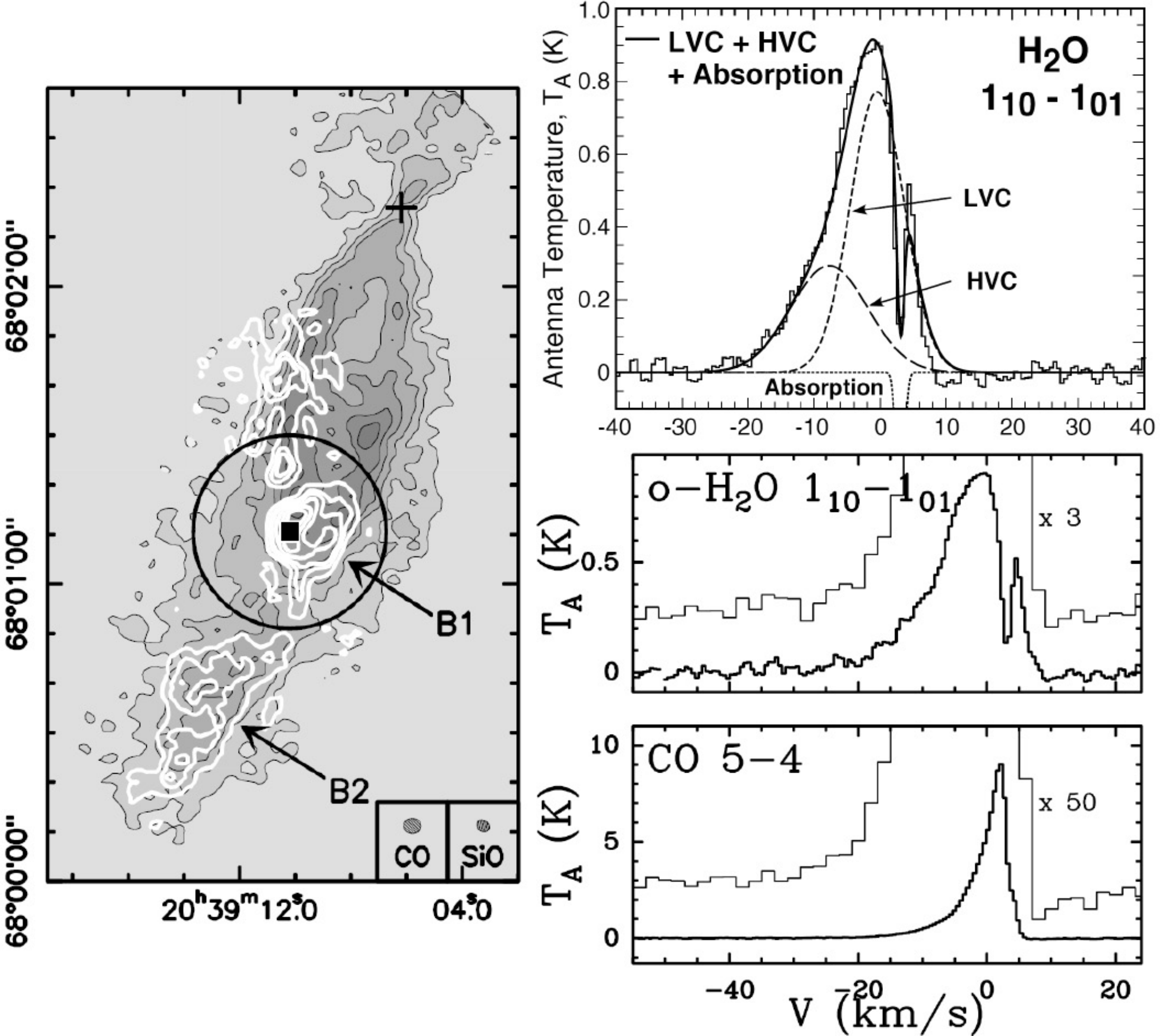}
\caption{{\em (left)}~Southern outflow lobe of L1157  in CO 1-0 (greyscale and
black contours) and in SiO 2-1 (white contours) as observed at the PdBI (Gueth
et al. 1996,98). A black square marks the nominal position of bowshock L1157-B1
observed with HIFI. The HIFI main-beam is represented with a black circle. {\em
(right)}~Panel of the CO 5-4 (bottom) and \gwater\ (centre) line spectra
obtained in band 1b of HIFI. For both lines, we show (dashed) a magnified and
spectrally smoothed view of the emission. Intensities are expressed in units of
antenna temperature.({\em top})~\water\ spectrum with fitted low-velocity
component (LVC), high-velocity component (HVC), absorption feature and summed
fitted spectrum.}
 \label{hifi_survey}
\end{figure}

The CO 5-4 transition is detected with an intensity  of $9\K$ ($\rm T_A$) and a
linewidth of $5\kms$. We notice a weak  absorption feature in the line profile
in the redshifted gas over a wide velocity range, which may partly arise from
cloud contamination in the reference position. The intensity in the blue wing
of the CO line differs by as much as 20\% between both polarizations. This
effect is not observed towards the \water\ line. Its origin is not understood
at the moment.

The fundamental \gwater\ line is detected with an intensity  of $0.9\K$ ($\rm
T_A$) at the peak. It is characterized by a broad linewidth  $\approx 15\kms$.
The line displays an absorption dip at $\rm v_{\rm lsr}= +2.9\kms$ and a broad
redshifted wing extending up to $+8\kms$. The broad linewidth of the \water\
spectrum could be fit with three Gaussian velocity components, a low-velocity,
a high-velocity, and an absorption component. The low (high) velocity component
peaks at $v_{\rm lsr}= -0.58\kms$ ($-7.86\kms$); the linewidth and peak
intensity derived from the fit are $9.56\kms$ and $0.77\K$ ($13.72\kms$ and
$0.29\K$), respectively. The absorption component was fit by a narrow line
Gaussian ($\Delta V= 1.38\kms$) of amplitude $-0.48\K$ centered at $v_{\rm
lsr}=+2.9\kms$. The fit of the individual components and the resulting fit to
the water spectrum is displayed in Fig.~1.

Overall, the \water\ and CO emission are detected in the same velocity range.
The high sensitivity of the HIFI observations permits the detection of emission
from the entrained gas up to $v_{\rm lsr}= -30\kms$, i.e. about $10\kms$ higher
than was previously known from ground based observations. However, line
profiles differ noticeably  and the ratio of the $\rm \water / CO 5-4$ line
intensities increases with increasing velocities from about 0.2 in the ambient
gas up to 0.9 at $\rm v_{\rm lsr}= -25\kms$ (Fig.~3).

All the other molecular tracers detected in the HIFI band show a pronounced
break in the line profile at $\rm v_{\rm lsr}\approx -7.2\kms$ (Codella et al.
this volume). This is also observed in the CO 6-5 spectrum of B1 obtained by us
at the CSO, as part of complementary observations to help analyse  the CHESS
data. The maps of the whole southern lobe of L1157 in CO 3-2 and 6-5 obtained
at the CSO in June 2009, with $24\arcsec$ and $14.5\arcsec$ respectively, will
be discussed in detail in a forthcoming paper (Lefloch et al. 2010, in prep.).

Below, we define the region with $\rm v_{\rm lsr} <-7.25\kms$  as the
high-velocity component, hereafter HVC (see also Codella et al), and the region
with $\rm v_{\rm lsr} > -7.25\kms$) as the low-velocity component (LVC). As we
discuss below, these two velocity components are characterized by different
spatial extents and excitation conditions.

\section{Discussion}

\begin{table*}
\caption{Observed and fitted parameters of the \water\ and CO 5-4 lines and
prediction for the $179\mum$ \water\ line flux. The CO fluxes in the HVC and
LVC are integrated in the velocity intervals [-30;-7.25] and [-7.25; +11.0],
respectively. The \water\ fluxes are derived from a multiple Gaussian fit to
the line profile. Intensities are expressed in units of antenna temperature
($T_{A}$) $\K\kms$.}
\begin{center}
\begin{tabular}{lcccccccccc
}
\hline
  &  $\water$~557GHz &  CO 5-4 & CO 6-5$^1$ & CO 3-2$^1$ & Size &
N(CO)$^2$ & n($\htwo$)$^2$ & T$^2$ & X($\water$)$^3$ & F($179\mum$)   \\
 &   ($\K\kms$) & ($\K\kms$) & ($\K\kms$) &  $\K\kms$ & ($\arcsec$) & ($\cmmd$)& ($\cmmt$) &
($\K$) & ($\cmmd$) & (W$\cmmd$) \\  \hline
LVC & 7.83 & 45.4 & 29.2 & 40.4 & 25 & 8.0(16) & (1.0-3.0)(5)  & 100 & 0.8(-6) &  4.2(-20)  \\
HVC & 4.28 & 3.98 & 1.34 & 3.11 & 7& 5.0(16) & (1.0-3.0)(4) & 400 & 0.8(-4) &  7.1(-20) \\
\hline
\end{tabular}
\end{center}
$^1$~From CSO observations smoothed down to the resolution of the HIFI observations.\\
$^2$~Determined from LVG analysis of the CO emission.\\
$^3$~From comparison of LVG-derived N($\rm o-\water$) with N(CO), assuming a
water OPR of 3 and an abundance $\rm [CO]/[$\htwo$] = 10^{-4}$.
\end{table*}

\subsection{Origin of the emission}

\begin{figure}
\includegraphics[width=\columnwidth]{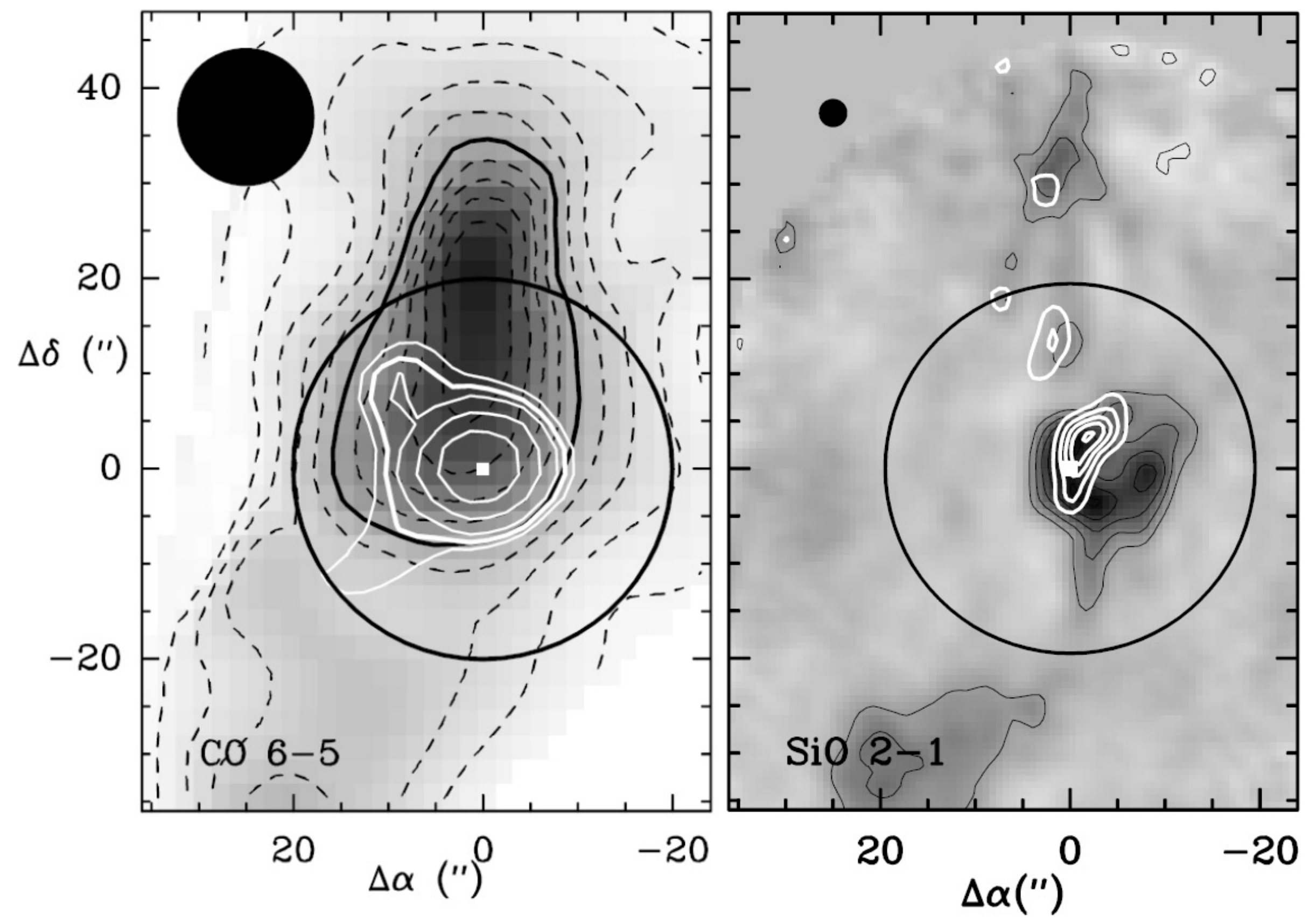}
\caption{{\em (left)}~Velocity-integrated CO 6-5 emission maps
 of the low- and high-velocity components. LVC (HVC) emission is represented
 in greyscale and thin dashed contours (white contours); first contour and contour
 interval are $3\sigma$ and $1\sigma$ ($10\%$ of the peak flux), respectively,
Contours at half-power are drawn in thick.
 {\em (right)}~Same for SiO 2-1 observed at the PdbI. The HIFI main-beam is represented
  with a black circle.}
\label{hifi_survey}
\end{figure}

Due to its relatively high energy above the ground state ($E_{up}= 116\K$) the
CO~6-5 transition is a good probe of the warm regions where \water\ can
evaporate from grain mantles and be released in the gas phase. SiO 2-1,
observed at the PdBI at $2.5\arcsec$ resolution  is a particularly good tracer
of shocks strong enough to release refractory elements in the gas phase,
because it is usually undetected in the cold, quiescent molecular gas.

The overall SiO emission is strongly peaked at the position of B1, which
appears as a region of $\approx 15\arcsec$ size located at the apex of the
cavity. Interferometric maps of the southern lobe (Figs.~1-2) reveal extended
emission along the eastern wall of the cavity (the low-velocity wing of the
bow) and downstream of B1, at velocities close to systemic, both blue and red.
By comparing these data with IRAM 30m observations (Bachiller et al. 2001), we
checked that unlike the HVC, a fraction of the flux emitted in the LVC is
actually missed in the PdBI data, which is direct evidence for extended
emission. This is consistent with the CO 6-5 data (Fig.~2). The low-velocity
gas emission is located in the wake of B1, reaching $\sim 40\arcsec$ North from
the apex. The area of the LVC amounts to $\approx 1/3$ of the HIFI beam (see
Fig.~2). Interestingly, the PACS map of the $\water$ $\rm 179\mum$ line reveals
large-scale emission, spatially coinciding with SiO~2-1 emission in the outflow
(Nisini et al. 2010).

In any case, there is definitely much less molecular gas emission associated
with the western wall of the cavity (Benedettini et al. 2007). We therefore
expect an asymmetry in the \water\ spatial distribution, similar to that
observed in many other tracers such as CS or HCN, as shown by the PACS map of
the $179\mum$ $\water$ line (Nisini et al. 2010).

We note an excellent agreement between the \water\ and the low-excitation
SiO~2-1 line profiles (Fig.~3) in the high-velocity range, with a constant SiO
2-1 / \water\ line ratio $\approx 0.8$ between -20 and $-7\kms$.  This has
important implications. First, this constant ratio in the range of the HVC
suggests that both emissions most likely arise from the same region and that
the emissions are optically thick. In that case, the low intensities measured
in the high-velocity component (a few tenths of K) point to a small size
extent. This is direct evidence that the \water\ emission detected fills only
partly the HIFI beam.  Indeed, the bulk of the SiO HVC originates from a small
region of $4\arcsec \times 12\arcsec$ in B1 (Fig.~2), corresponding to a
filling factor $\rm ff \sim 0.03$ in the HIFI main-beam.
Second, if silicon comes from grain erosion, the SiO profile is predicted to be
much narrower than \water\ because it takes a long time for Si to oxidize into
SiO, so SiO comes only from the cold postshock, as discussed by Gusdorf et al.
(2008b). The similarity of the SiO and \water\ line profiles suggests that SiO
forms more extensively in the shock than predicted by oxidation of sputtered Si
atoms. As it can be released in the gas phase even at low velocities in the
shock, SiO is present in the gas phase over the full width of the shock wave.

\begin{figure}
\includegraphics[width=\columnwidth]{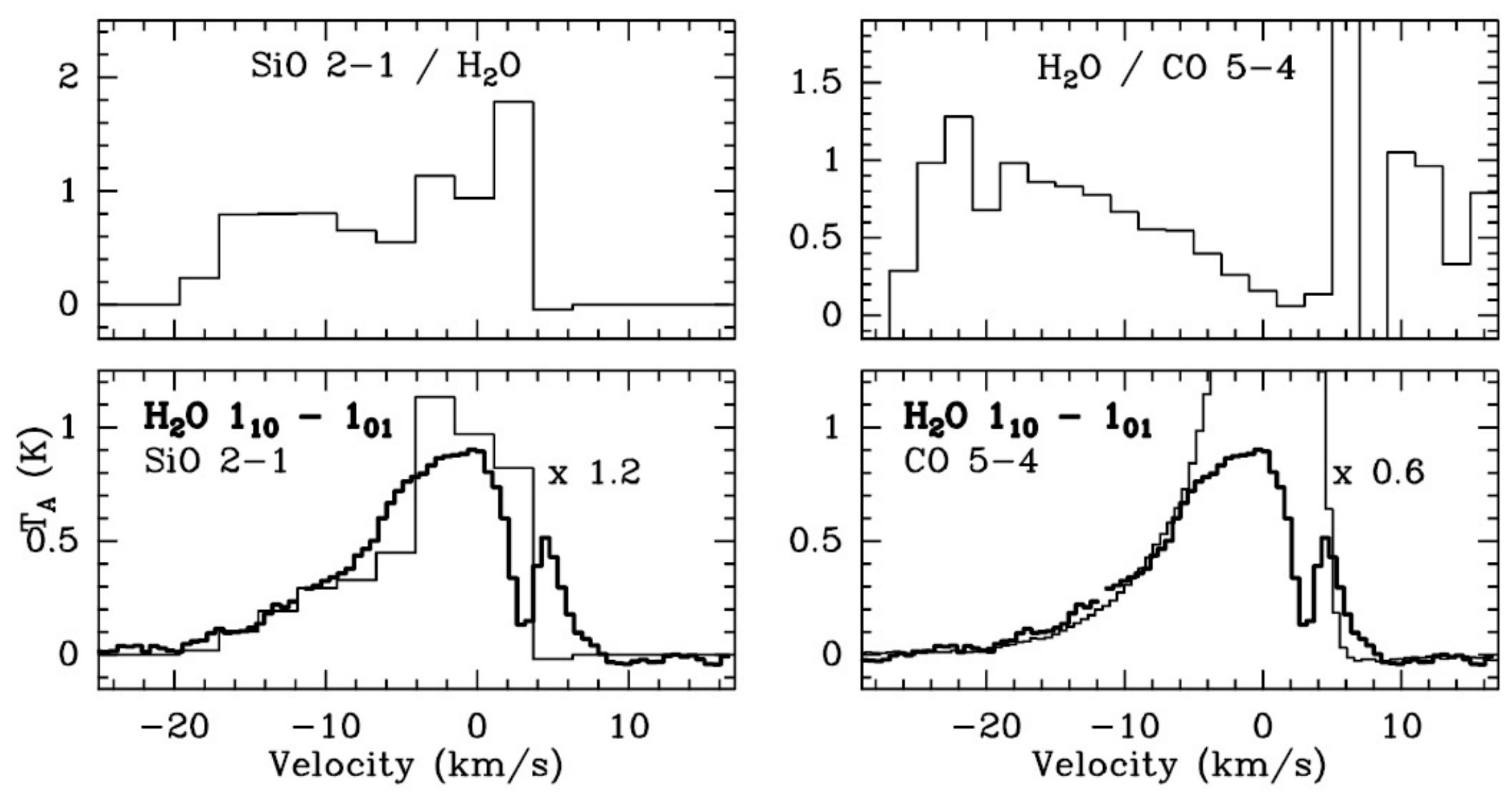}
\caption{({\em bottom left})~Comparison of the \gwater\ line profile with the
SiO 2-1 emission observed at the PdBI, averaged over the HIFI beam. ({\em top
left}) Variations of the SiO 2-1 / \water\ line ratio as a function of
velocity. {\em (bottom right)}~Comparison of the \gwater\ and CO 5-4 line
profiles.  {\em (top right)}~Variations of the \water\ /CO 5-4 line ratio as a
function of velocity, smoothed to a resolution of $2\kms$.}
 \label{hifi_survey}
\end{figure}

In summary, we find strong observational evidence that the emission from the
HVC and LVC arises from regions of different physical extent. The size of the
HVC appears definitely much lower than the LVC ($ff\simeq 0.03$ and 0.3,
respectively). It is true however that the present determinations are
uncertain. HIFI observations of the high-excitation lines of CO and \water\
will make it possible to better establish appropriate filling factors for these
components.

\begin{figure}
\includegraphics[width=\columnwidth]{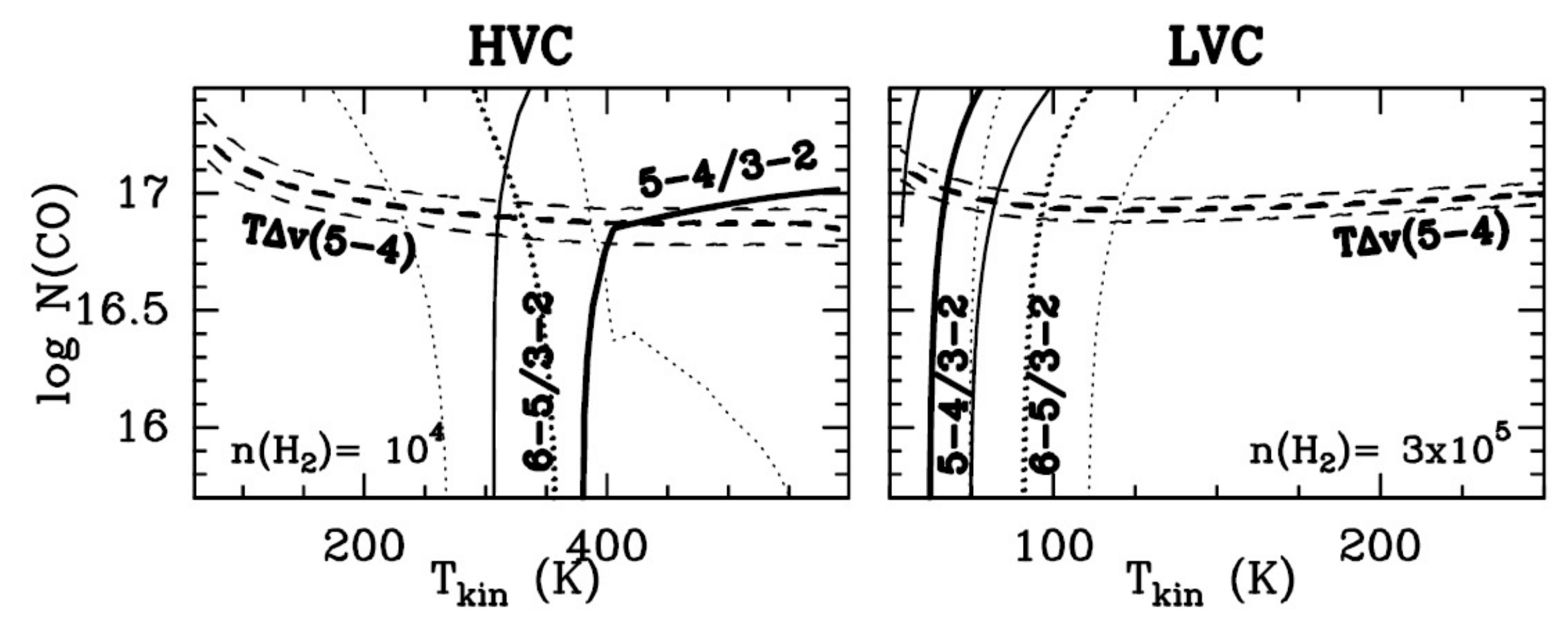}
\caption{Best-fit solution to the LVG modelling of CO line temperatures  in the
3-2, 5-4 and 6-5 transitions for both HVC (left) and LVC (right) components,
respectively. The contour of the observed CO 5-4 integrated line area ($\rm
T\Delta v $), corrected for main-beam dilution, is drawn in a dashed line, as
well as the 5-4/3-2 and 6-5/3-2 line ratios (solid and dotted lines,
respectively). Thin lines delineate the uncertainties in the observed values.}
\end{figure}

\subsection{Physical conditions}

We first estimated the physical conditions from the emission detected in the
CO~3-2, 5-4, and 6-5 transitions both in HVC and LVC . We modelled each
velocity component as a simple uniform slab, adopting the size (filling factor)
estimated above. Calculations were done in the large-velocity gradient
approach, using the CO collisional coefficients determined by Flower (2001) for
ortho-\htwo\ collisions in the range $5\K$-$400\K$. For temperatures beyond
$400\K$, the collisional coefficients were extrapolated adopting a temperature
dependence of $\sqrt{T/400\K}$. Both components appear to have  the same gas
column density $\rm N(CO) \simeq 10^{17}\cmmd$ (Fig.~4). We found the
high-velocity gas component to be unambiguously associated with hot gas ($T
> 350\K$)  of moderate density $\approx 3.0\times 10^4\cmmt$, whereas  the
low-velocity component arises from  gas at a lower temperature ($100\K$) and
higher density ($\sim 3.0\times 10^5\cmmt$). The temperature estimated for the
extended component agree reasonably well with other determinations from $\rm
NH_3$ and $\rm CH_3CN$ (Tafalla \& Bachiller 1995; Codella et al. 2009).

With the physical conditions derived from the CO analysis, we modelled the
integrated intensity and the line profile of the \gwater\ transition as well as
the reported PACS-measured $179\mum$ \water\ line intensity ($\sim 10^{-19}\rm
W \cmmd$, Nisini et al. 2010) to compute the total ortho water abundance in
each velocity component. We used a radiative transfer code in the
large-velocity gradient approach (and slab geometry) detailed in Melnick et al.
(2008), taking into account an ortho to para ratio (OPR) of 1.2 for $\htwo$, as
derived from {\em Spitzer} (Neufeld et al., 2009). Here, we assume the
absorption component at $+2.9\km$ is due to foreground gas unrelated to
L1157-B1. Together, the two components of the \gwater\ line produce a total
\water\ $2_{12}-1_{01}$ $179\mum$ line flux of $\rm 1.1\times 10^{-19}W\cmmd$.
For the temperature range derived from our CO analysis, we estimated
ortho-$\water$ column densities of $(4.0-5.0)\times 10^{14}\cmmd$ and
$(2.5-3.0)\times 10^{16}\cmmd$ for the LVC and the HVC, respectively. Assuming
an OPR of 3, we derived the \water\ abundance  from comparison with the gas
column densities estimated from CO (see Table~1). We obtained an abundance
ratio $\rm [\water]/[CO] \simeq 0.8$ in the high-velocity gas, which is
consistent with previous results from ODIN (Benedettini et al. 2002) and agrees
reasonably well with the predictions of steady-state C-shock models for this
set of physical parameters (shock velocity $V_{s}\simeq 20\kms$, pre-shock
density $n(\htwo)= 5\times 10^3\cmmt$; Gusdorf et al. 2008a,b).

An interesting prediction of our model is that the HVC contribution to the
$179\mum$ flux dominates over the LVC contribution (see last column in
Table~1). The higher temperature of this component drives the neutral-neutral
reactions that efficiently form $\water$, and the higher shock velocity can
more efficiently remove water from grain mantles (see Melnick et al. 2008),
resulting in the much greater ortho-\water\ column density than in the LVC.
Comparison with NH$_3$ also suggests that the water production in the HVC is
strongly dominated by high-temperature reactions (see Codella et al.). The
higher ortho-\water\ column density is what produces the higher $179\mum$ line
flux from this component. Consistent results are obtained by Nisini et al.
(2010) based on a $179\mum$ PACS map and previous ODIN and SWAS observations of
the \gwater\ line, assuming one single physical component dominates the water
line emission in the HIFI beam. Follow-up observations of the higher-excitation
lines of CO and \water\ with HIFI will help us constrain more accurately the
physical conditions of each velocity component (density, temperature) and more
generally in the shock.



\acknowledgements
HIFI has been designed and built by a consortium of institutes and university
departments from across Europe, Canada and the United States under the
leadership of SRON Netherlands Institute for Space Research, Groningen, The
Netherlands and with major contributions from Germany, France and the US.
Consortium members are: Canada: CSA, U.Waterloo; France: CESR, LAB, LERMA,
IRAM; Germany: KOSMA, MPIfR, MPS; Ireland, NUI Maynooth; Italy: ASI, IFSI-INAF,
Osservatorio Astrofisico di Arcetri- INAF; Netherlands: SRON, TUD; Poland:
CAMK, CBK; Spain: Observatorio Astron\'{o}mico Nacional (IGN), Centro de
Astrobiolog\'{\i}a (CSIC-INTA). Sweden: Chalmers University of Technology - MC2, RSS
\& GARD; Onsala Space Observatory; Swedish National Space Board, Stockholm
University - Stockholm Observatory; Switzerland: ETH Zurich, FHNW; USA:
Caltech, JPL, NHSC. HIPE is a joint development by the {\em Herschel} Science
Ground Segment Consortium, consisting of ESA, the NASA {\em Herschel} Science
Center, and the HIFI, PACS and SPIRE consortia.
\def\baselinestretch{0.8}

{}

\end{document}